\documentclass[useAMS,usenatbib,usegraphicx]{mn2e}

\title[Possible dual radio-emitting nucleus in SDSS J1425+3231]{Two in one? A possible dual radio-emitting nucleus in the quasar SDSS J1425+3231}

\author[S. Frey, Z. Paragi, T. An and K.\'E. Gab\'anyi]{S. Frey$^{1}$\thanks{E-mail:
frey@sgo.fomi.hu},
Z. Paragi$^{2}$, T. An$^{3,4,5}$ and K.\'E. Gab\'anyi$^{6,1}$\\
$^{1}$F\"OMI Satellite Geodetic Observatory, P.O. Box 585, H-1592 Budapest, Hungary\\
$^{2}$Joint Institute for VLBI in Europe, Postbus 2, 7990 AA Dwingeloo, The Netherlands\\
$^{3}$Shanghai Astronomical Observatory, Chinese Academy of Sciences, 80 Nandan Road, 200030 Shanghai, P.R. China\\
$^{4}$Netherlands Institute for Radio Astronomy (ASTRON), Postbus 2, 7990 AA Dwingeloo, The Netherlands\\
$^{5}$Key Laboratory of Radio Astronomy, Chinese Academy of Sciences, P.R. China\\
$^{6}$Konkoly Observatory, MTA Research Centre for Astronomy and Earth Sciences, P.O. Box 67, H-1525 Budapest, Hungary}

\begin{document}

\date{Accepted 2012 June 9. Received 2012 May 31; in original form 2012 March 22}

\pagerange{\pageref{firstpage}--\pageref{lastpage}} \pubyear{2012}

\maketitle

\label{firstpage}

\begin{abstract}
The radio-emitting quasar SDSS~J1425+3231 ($z$=0.478) was recently found to have double-peaked narrow [O III] optical emission lines. Based on the analysis of the optical spectrum, \citet{Peng11} suggested that this object harbours a dual active galactic nucleus (AGN) system, with two supermassive black holes (SMBHs) separated on the kpc scale. SMBH pairs should be ubiquitous according to hierarchical galaxy formation scenarios in which the host galaxies and their central black holes grow together via interactions and eventual mergers. Yet the number of presently-confirmed dual SMBHs on kpc or smaller scales remains small. A possible way to obtain direct observational evidence for duality is to conduct high-­resolution radio interferometric measurements, provided that both AGN are in an evolutionary phase when some activity is going on in the radio. We used the technique of Very Long Baseline Interferometry (VLBI) to image SDSS~J1425+3231. Observations made with the European VLBI Network (EVN) at 1.7~GHz and 5~GHz frequencies in 2011 revealed compact radio emission at sub-mJy flux density levels from two components with a projected linear separation of $\sim$2.6 kpc. These two components support the possibility of a dual AGN system. The weaker component remained undetected at 5~GHz, due to its steep radio spectrum. Further study will be necessary to securely rule out a jet--shock interpretation of the less dominant compact radio source. Assuming the dual AGN interpretation, we discuss black hole masses, luminosities, and accretion rates of the two components, using available X-ray, optical, and radio data. While high-resolution radio interferometric imaging is not an efficient technique to search blindly for dual AGN, it is an invaluable tool to confirm the existence of selected candidates.

\end{abstract}


\section{Introduction}

It is widely accepted that most major galaxies harbour supermassive black holes (SMBHs) in their nuclei. Hierarchical structure formation models naturally involve interactions and mergers in which the host galaxies and their central SMBHs grow together \cite[e.g.][]{Kauf00}. In this scenario, we expect to see some dual SMBH systems as snapshots of the corresponding evolutionary phases of the merger process. Accretion of the surrounding gas onto the SMBHs may give rise to activity and make the objects prominent across the whole electromagnetic spectrum. If this activity is going on at both components of the SMBH pair during the same period of time, we may observe dual Active Galactic Nucleus (AGN) systems. Recent high-resolution smoothed particle hydrodynamical simulations \citep{Wass12} suggest that simultaneous AGN activity is mostly expected at the late phases of mergers, at or below $\sim$1--10 kpc-scale separations. However, due to the high spatial resolution required, it is difficult to identify such systems in practice. 
Indirect signatures of duality are often inconclusive and require other supporting observational evidence.
Therefore the models of dual SMBH evolution are still poorly constrained by observations. In the last couple of years, the search for dual AGN has become increasingly popular. The numerous theroetical and observational results are reviewed by e.g. \citet{Komo06}, \citet{Dott12}, and \citet{Popo12}.   
 
The presence of double-peaked narrow optical emission lines are thought to indirectly indicate AGN pairs. The large spectroscopic data base of the Sloan Digital Sky Survey (SDSS) made it possible to select many extragalactic objects with double-peaked [O III] emission line profiles. These lines may originate from distinct narrow-line regions (NLRs) of two gravitationally bound SMBHs, with ``intermediate'' ($\sim$1--10~kpc) separations \citep[e.g.][]{Smit10}. Such objects should have spatially distinct NLRs, and the typical orbital velocities in the host galaxies are of the order of 100~km~s$^{-1}$. After examining more than 20\,000 spectra, \citet{Smit10} estimated that about 1 per cent of broad-line (Type 1) SDSS quasars at 0.1$<$$z$$<$0.7 have double-peaked narrow-line profiles. However, there exist viable alternative explanations for the double-peaked narrow lines related to peculiar kinematics and jet--cloud interaction in a single NLR. Detailed studies are needed to confirm or reject the dual-AGN scenario for each individual object. Using near-infrared imaging and optical slit spectroscopy of Type 2 AGN with double-peaked [O III] emission lines, \citet{Shen11} found that only 10 per cent of their targets are definitely dual AGN at kpc-scale projected separations. \citet{Rosa11} studied a sample of 12 dual AGN candidates with near-infrared laser guide star adaptive optics imaging, and estimated that $\sim$0.3--0.65 per cent of the SDSS quasars host dual accreting black holes separated on kpc scales. This agrees well with the order-of-magnitude estimate of \cite{Fu11a} who claim that the kpc-scale binary fraction of AGN is $<$0.3 per cent.

According to \citet{Dott12}, only about 20 dual AGN systems with separations of $\sim$10~pc to $\sim$10~kpc were known at the time of writing their review. Even if the number has increased in the meantime, it probably does not exceed a few dozen. Radio-emitting pairs are at least as rare. 
The radio galaxy 3C\,75 (NGC\,1128) is a dual source with prominent arcsecond-scale radio jets \citep{Owen85}. The X-ray counterparts of the two SMBHs with a projected separation of 7~kpc are resolved in the {\it Chandra} image of \citet{Huds06}.   
In the radio galaxy 0402+379, two bright flat-spectrum cores separated by 7.3~pc were found with Very Long Baseline Interferometry (VLBI) observations by \citet{Rodr06}. Remarkably, an extensive systematic study of \cite{Burk11} failed to identify any flat-spectrum radio pair other than 0402+379 among the most frequently observed 3114 luminous radio AGN for which archival VLBI data at two or more different frequencies were available. However, when looking for dual AGN in the radio, we don't have to restrict ourselves to bright and/or flat-spectrum radio sources. Recently \citet{Ting11} examined a sample of 11 radio AGN showing double-peaked [O III] optical emission lines with the US National Radio Astronomy Observatory (NRAO) Very Long Baseline Array (VLBA) at 1.4~GHz. They did not find any dual source above their $\sim$1.5 mJy flux density detection threshold. On the other hand, there are cases when AGN pairs are successfully found in the radio among weaker radio sources, at even fainter (sub-mJy) flux density levels. \citet{Bond10} detected two comapct VLBI cores in the radio-quiet candidate binary black hole system SDSS J1536+044. The measured flux densities at 5~GHz were 0.72 and 0.24~mJy, the separation is $\sim$5~kpc. \citet{Fu11b} used the Expanded VLA (EVLA) at three different frequencies (1.4, 5, and 8.5~GHz) to confirm that SDSS J1502+1115, a double-peaked [O III] emission-line quasar with two components in its near-infrared adaptive optics image is also a dual radio source. The projected linear separation of this pair is 7.4~kpc, the source flux densities are at the mJy or sub-mJy level.

The subject of our present study, the quasar SDSS J142507.32+323137.4 (SDSS J1425+3231, or J1425+3231 in short) 
at the redshift $z$=0.478 was identified by \citet{Peng11} as having double-peaked narrow [O III] 4959 and 5007~\AA\ optical emission lines in its SDSS spectrum. The authors modelled the line profiles with three Gaussian components: one for the blue narrow-line component (i.e. approaching with respect to the systemic redshift), one for the red narrow (i.e. receding) component, and one for the underlying broad wing whose peak is close to the blueshifted narrow line. The blue and red components are separated by $\sim$500~km~s$^{-1}$ in velocity. \citet{Peng11} proposed that J1425+3231 is a dual AGN system, with kpc-scale separation. The more massive of the two AGN corresponds to the blue emission line component, which is about twice as broad as the red one. 
The broad blueshifted wing arises in its intermediate-line region. \citet{Peng11} estimate $\sim$$10^8$~M$_\odot$ for the mass of the primary SMBH.
The suggested secondary black hole is smaller ($\sim$$10^6$\,M$_\odot$) and could be a Type 2 AGN (i.e. seen close to the plane of the obscuring material surrounding the central accreting black hole). 
The absence of broad emission lines from the secondary AGN may be due to obscuration by a dusty torus, or the SDSS spectrum is insufficient to separate the broad lines originating from the two nearby AGN. 

The optical images taken from the SDSS Data Release 7 archive server\footnote{\tt {http://cas.sdss.org/dr7/en/}} \citep{Abaz09} show a point-like object with no indication of any extension at arcsecond angular scale. At 1.4 GHz, the source is unresolved ($<$$5\arcsec$) in the Very Large Array (VLA) Faint Images of the Radio Sky at Twenty-centimeters (FIRST) survey\footnote{\tt{http://sundog.stsci.edu}} \citep{Whit97}, with $S_{\rm 1.4}$=3.28~mJy integrated flux density. Its compact radio emission and dual AGN candidature make J1425+3231 potentially interesting for high-resolution radio interferometric imaging.

\begin{figure*}
\centering
  \includegraphics[bb=50 177 570 667, height=60mm, angle=0, clip=]{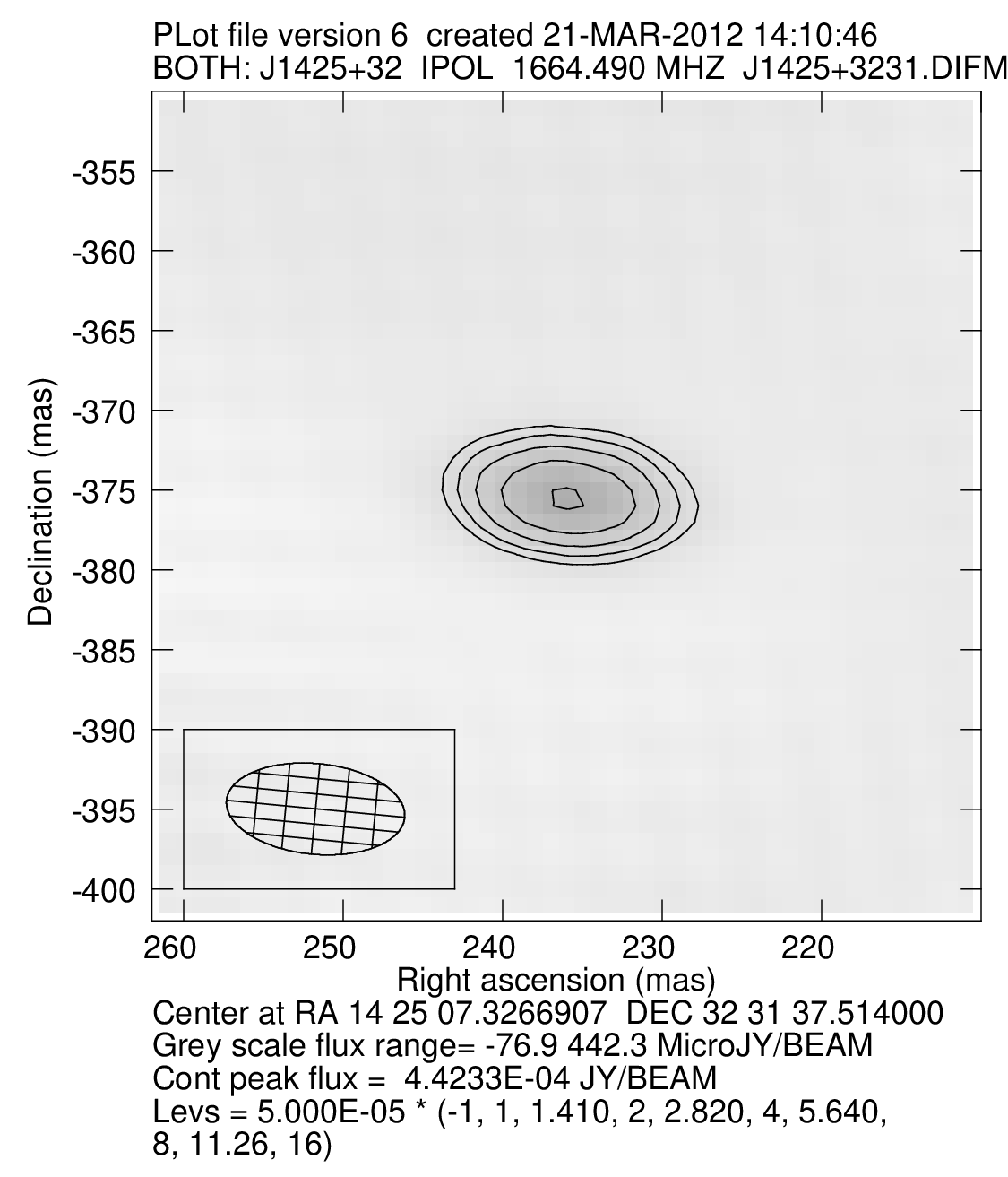}
  \includegraphics[bb=63 115 553 713,  height=60mm, angle=0, clip=]{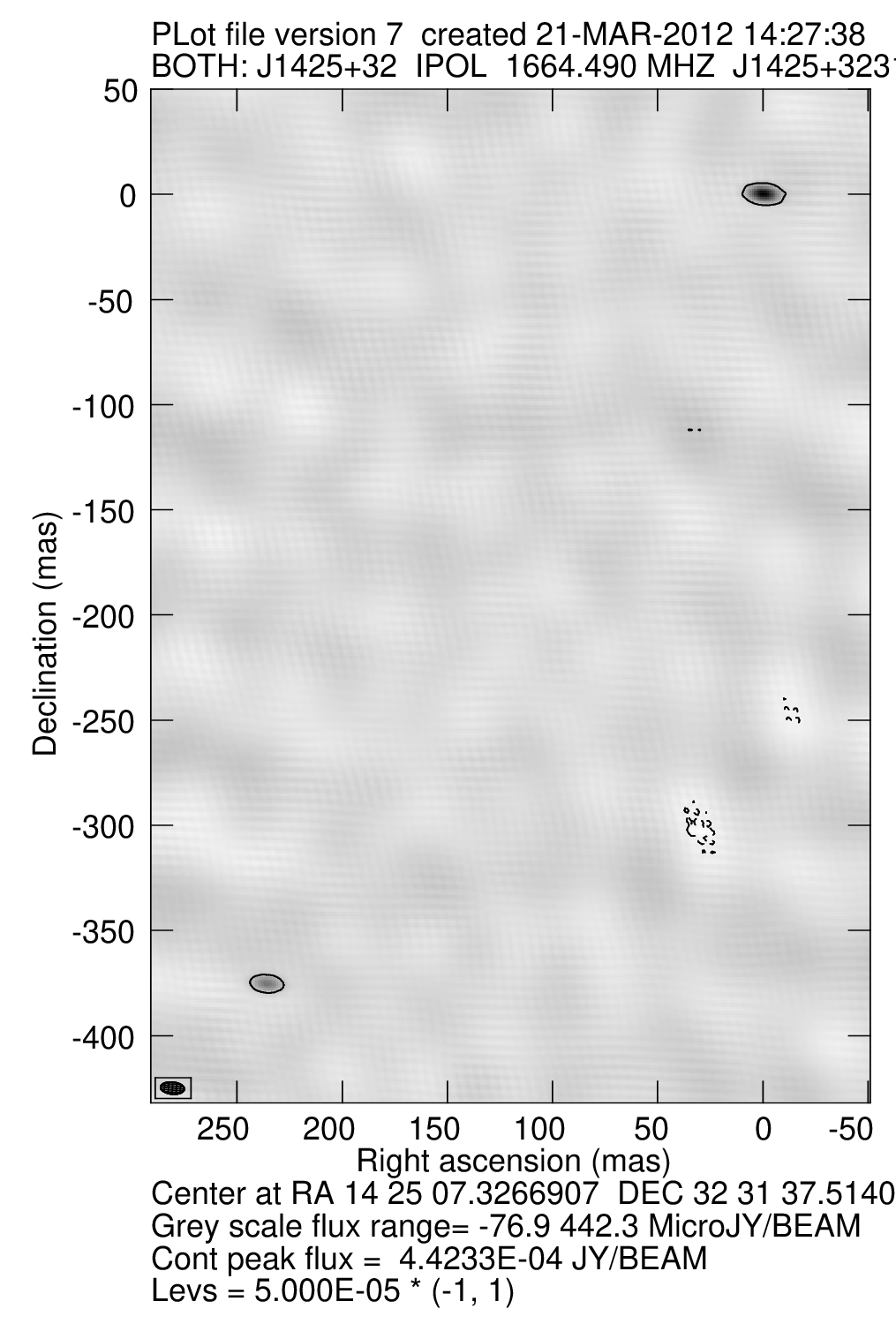}
  \includegraphics[bb=55 173 575 672, height=60mm, angle=0, clip=]{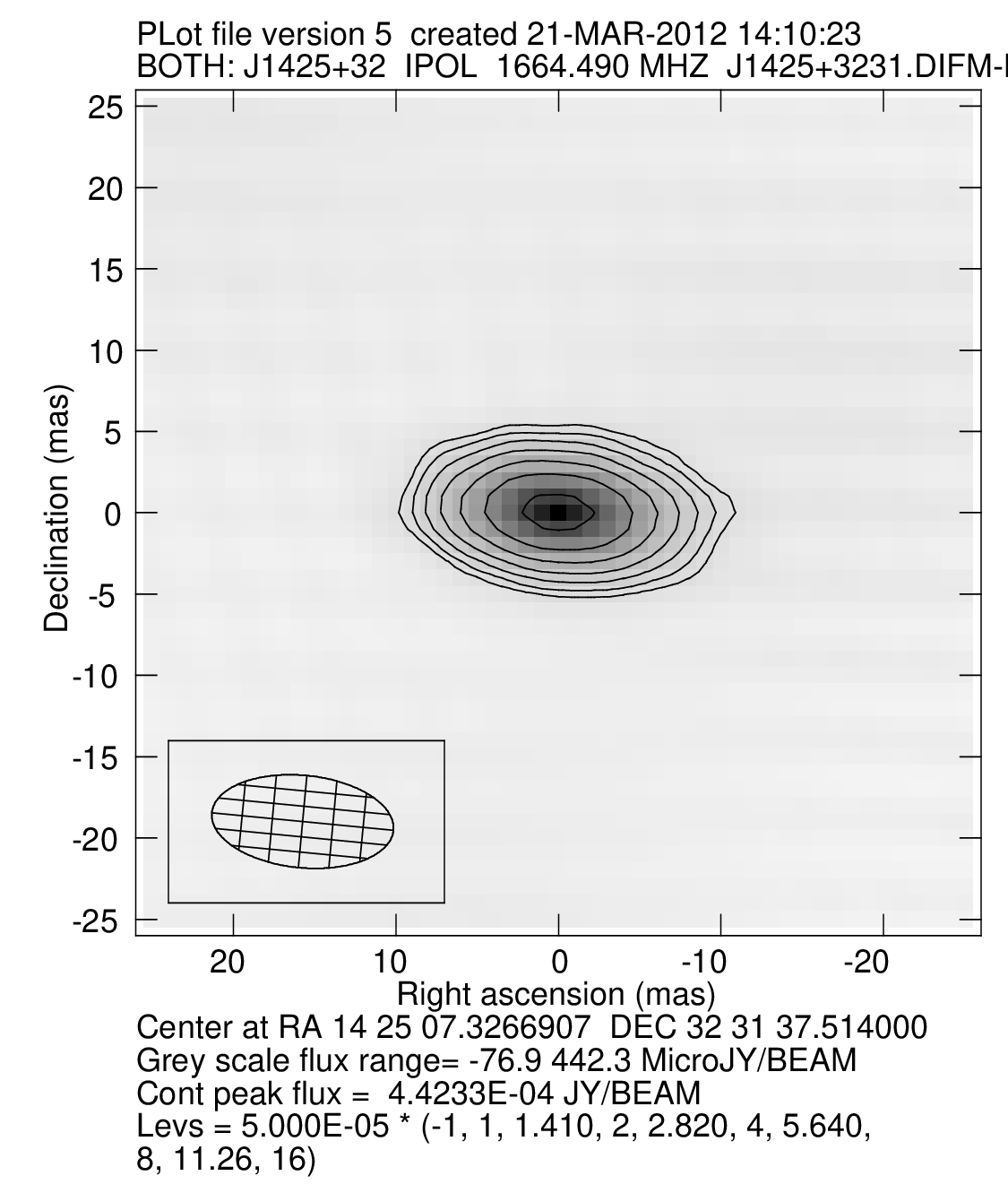}
  \caption{
The naturally-weighted 1.7-GHz image of the quasar J1425+3231 from the EVN experiment RSF04 on 2011 January 26 {\it (middle)}. Close-up images of the two individual components are shown separately. The lowest negative and positive contours are drawn at $\pm50$~$\mu$Jy~beam$^{-1}$ (3$\sigma$). Further positive contour levels (displayed only in the close-up images) increase by a factor of $\sqrt2$. The peak brightness for the north-western component {\it (right)} is 442~$\mu$Jy~beam$^{-1}$, and for the south-eastern component {\it (left)} is 205~$\mu$Jy~beam$^{-1}$. The Gaussian restoring beam is 11.2~mas $\times$ 5.7~mas with major axis position angle $85\degr$. The restoring beam (full width at half maximum, FWHM) is indicated with ellipses in the lower-left corners.}
  \label{Lband-image}
\end{figure*}

Here we report on our VLBI imaging observations of J1425+3231 made with the European VLBI Network (EVN) at 1.7~GHz and 5~GHz frequencies,
at three different epochs in 2011. Our aim was to verify the prediction of \citet{Peng11} about the dual nature of this AGN, and to test how the EVN could possibly be used in the future for investigating a larger sample of candidate radio AGN pairs identified by their double-peaked [O III] emission lines. To calculate linear sizes and luminosities, we assume a flat cosmological model with $H_{\rm 0}$=70~km~s$^{-1}$~Mpc$^{-1}$, $\Omega_{\rm m}$=0.3, and $\Omega_{\Lambda}=$0.7. In this model, $1\arcsec$ angular size corresponds to 5.957~kpc projected linear size at $z$=0.478, and the luminosity distance of J1425+3231 is $D_{\rm L}$=2684.2~Mpc \citep{Wrig06}.

\section{VLBI observations and data reduction}

We initiated short exploratory EVN observations of J1425+3231 to check whether the radio source contains two compact components that might indicate a dual AGN. These 1.7-GHz observations used the e-VLBI mode \citep{Szom08} in which the signals from the remote radio telescopes are not recorded but streamed to the central data processor over optical fibre networks for real-time correlation. The experiment lasted for 2 h on 2011 January 26 (project code RSF04). The participating radio telescopes were Effelsberg (Germany), Jodrell Bank Lovell Telescope, Cambridge (UK), Medicina (Italy), Onsala (Sweden), Toru\'n (Poland), Hartebeesthoek (South Africa), Sheshan (China), and the phased array of the 14-element Westerbork Synthesis Radio Telescope (WSRT, the Netherlands). The maximum data transmission rate was 1024~Mbit~s$^{-1}$, which resulted in a total bandwidth of 128~MHz in both left and right circular polarizations, using 2-bit sampling.

J1425+3231, our weak target source, was observed in phase-reference mode. This helps to increase the total coherent integration time spent on the source and thus to improve the sensitivity of the observations. Phase-referencing is performed by regularly altering the pointing direction between the target source and a bright, compact, nearby reference source \citep[e.g.][]{Beas95}. We selected J1422+3223 as the phase calibrator from the VLBA Calibrator Survey list\footnote{\tt {http://www.vlba.nrao.edu/astro/calib/index.shtml}}. The target--reference angular separation is $0\fdg57$. The delay, delay rate, and phase solutions derived for the phase-reference calibrator were interpolated and applied to J1425+3231 within the cycle time of $\sim$6 min. The target source was observed for nearly 4.5 min in each cycle, resulting in a total of $\sim$80 min on-source integration time.

Since the exploratory 1.7-GHz observations clearly revealed that the target source is resolved into two distinct components separated by $\sim$440 milli-arcseconds (mas), we proposed a follow-up e-VLBI experiment at 5~GHz. The data at the higher frequency promised essential spectral information on the radio emission originating from the two components. The 5-GHz phase-referenced EVN observations were performed on 2011 October 18 and lasted for 6 h (project code EF023A). The radio telescopes at Effelsberg, Jodrell Bank Mk2, Medicina, Onsala, Toru\'n, Yebes (Spain), Sheshan, and the WSRT participated. With a similar setup as before, the total on-source time was 220 min.

We were also granted another 2-h experiment at the lower frequency (1.7~GHz) with the EVN, closer to the date of the 5-GHz observations. Data from this second observing epoch served to check whether the radio emisson of the components in J1425+3231 is variable over a period of $\sim$9 months. 
The repeated 1.7-GHz EVN experiment (project code EF023B) also lasted for 2 h but used disk-based recording at the following stations: Effelsberg, Jodrell Bank Lovell Telescope, Medicina, Onsala, Toru\'n, Svetloe, Zelenchukskaya, Badary (Russia), Sheshan, Urumqi (China), and the WSRT. The observing date was 2011 November 4. The correlation of the data from the disk-based and both e-VLBI experiments took place at the EVN MkIV Data Processor at the Joint Institute for VLBI in Europe (JIVE) in Dwingeloo, the Netherlands.

The NRAO Astronomical Image Processing System ({\sc AIPS}) was used for the data calibration in a standard way \citep[e.g.][]{Diam95}. The visibility amplitudes were calibrated using system temperatures and antenna gains measured at the telescope sites. Fringe-fitting was performed for the calibrator (J1422+3223) and fringe-finder sources (4C\,39.25, OQ\,208) using 3-min solution intervals. The data were then exported to the Caltech {\sc Difmap} package \citep{Shep94} for imaging. The conventional hybrid mapping procedure involving several iterations of CLEANing \citep{Hogb74} and phase (then amplitude) self-calibration resulted in the images and brightness distribution models for the calibrators. Overall antenna gain correction factors (typically $\sim$10 per cent or less) were determined and applied to the visibility amplitudes in {\sc AIPS}. Fringe-fitting was repeated for the phase-reference calibrator in {\sc AIPS}, now taking its CLEAN component model into account in order to compensate for residual phases resulting from its non-pointlike structure. The solutions obtained were interpolated and applied to the target source data. The calibrated and phase-referenced visibility data of J1425+3231 were also transferred to {\sc Difmap} for imaging. 

\begin{figure}
\centering
  \includegraphics[bb=55 173 570 672, height=60mm, angle=0, clip=]{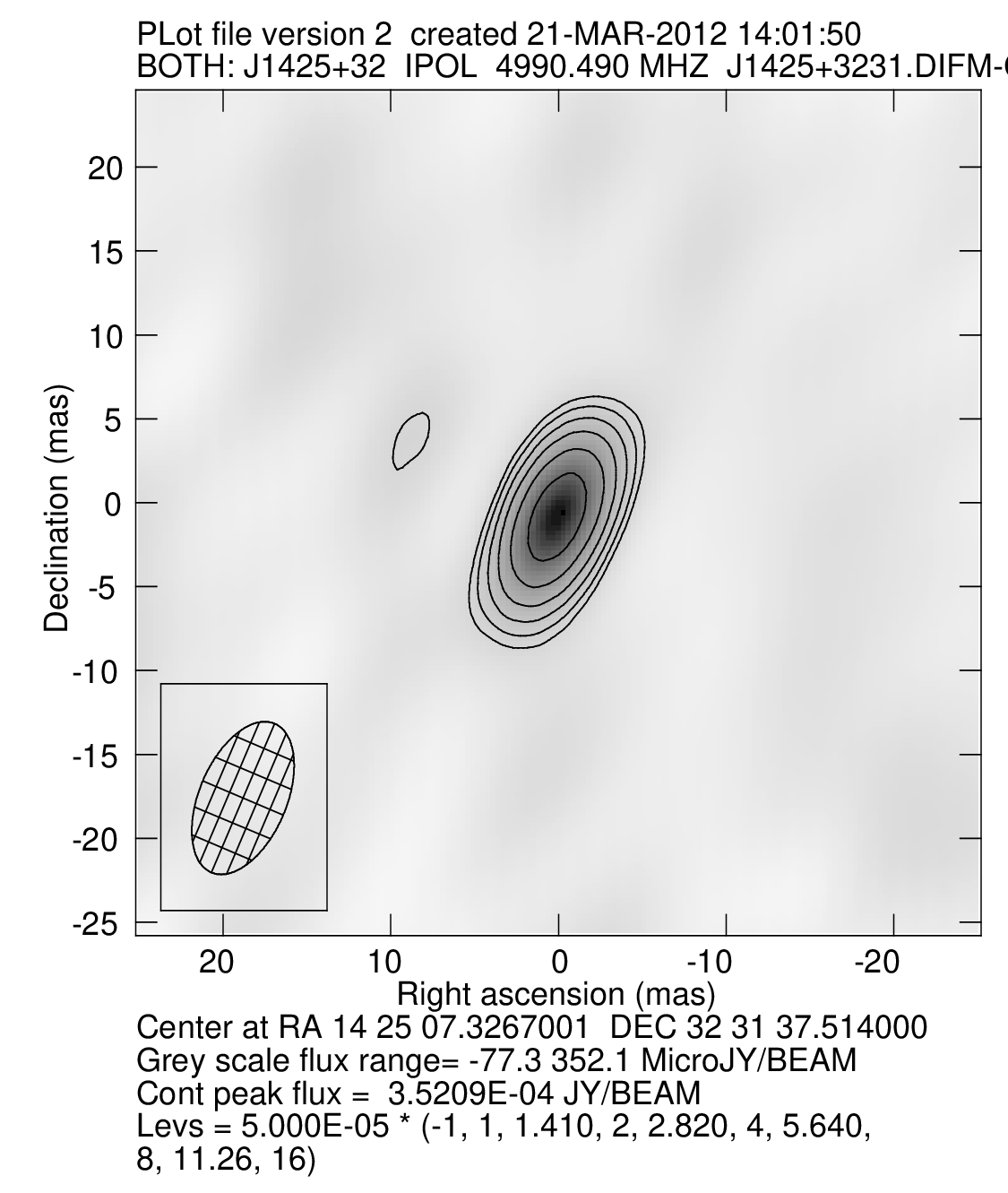}
  \caption{
The naturally-weighted 5-GHz EVN image of the brighter (north-western) component of the dual quasar J1425+3231. The lowest contours are drawn at $\pm50$~$\mu$Jy~beam$^{-1}$ (3$\sigma$). The positive contour levels increase by a factor of $\sqrt2$. The peak brightness is 352~$\mu$Jy~beam$^{-1}$. The Gaussian restoring beam is 9.6~mas $\times$ 5.2~mas with major axis position angle $-23\degr$.} 
  \label{Cband-image}
\end{figure}

The total intensity images at 1.7~GHz (Fig.~\ref{Lband-image}) and 5~GHz (Fig.~\ref{Cband-image}) were restored in {\sc Difmap} after fitting circular Gaussian brightness distribution models to the interferometric visibility data. Natural weighting was applied to achieve the lowest image noise. No self-calibration was attempted for the target source data. 

\section{The two radio components of J1425+3231}

There are two distinct compact radio sources seen in the 1.7-GHz image (Fig.~\ref{Lband-image}) made on 2011 January 26. 
At the second epoch (2011 November 4), the brighter (north-western, NW) source has also been detected. Its flux density and size agreed with the previous results within the errors. The detection of the fainter (south-eastern, SE) source proved difficult because of the elevated image noise level ($\sigma$=42~$\mu$Jy~beam$^{-1}$) we attribute to severe radio frequency interference apparent in the data, especially at the telescopes with the largest collecting area (Effelsberg and the WSRT). 
The non-detection of the fainter component at the 5$\sigma$ level is consistent with its brightness value measured in the first 1.7-GHz experiment (Fig.~\ref{Lband-image}), but the quality of the second-epoch data does not allow us to constrain the variability of this source.
     
At 5~GHz, only the brighter component is detected (Fig.~\ref{Cband-image}). In both the 1.7-GHz and 5-GHz images, the lowest contours are drawn at $\sim$$3\sigma$ image noise levels. The coordinates are relative to the position of the brightness peak at 5~GHz. The phase-referenced absolute equatorial coordinates for the NW source are right ascension $\alpha_{\rm NW}$=$14^{\rm h}25^{\rm m}07\fs32669$ and declination $\delta_{\rm NW}$=$32\degr31\arcmin37\farcs5133$ (J2000). These were determined from the 5~GHz data with the accuracy of 0.4~mas each, and are identical with the 1.7-GHz coordinates within the errors. The position of the SE source is $\alpha_{\rm SE}$=$14^{\rm h}25^{\rm m}07\fs34535$ and $\delta_{\rm SE}$=$32\degr31\arcmin37\farcs1384$, as determined from the 1.7-GHz data.
The angular separation of the two components is 443.0$\pm$0.8~mas, which corresponds to 2.639$\pm$0.005~kpc projected linear distance. 

The circular Gaussian brightness distribution models fitted in {\sc Difmap} at both frequencies allow us to characterise the sources. The model parameters and the derived brightness temperatures ($T_{\rm B}$) are listed in Table~\ref{modelfit}. The statistical errors are estimated according to \citet{Foma99}. Additional flux density calibration uncertainties are assumed as 5 per cent.

\begin{table*}
  \centering 
  \caption[]{Parameters of the fitted circular Gaussian models for the NW component {\it (top)} and the SE component {\it (bottom)} of J1425+3231, and the derived brightness temperatures, radio powers, and luminosities. In the case of the non-detection of the SE component at 5~GHz, we assume an unresolved VLBI source to obtain an upper limit of the flux density. }
  \label{modelfit}
\begin{tabular}{cccccccc}        
\hline                 
Flux density & Frequency & \multicolumn{2}{c}{Relative position } & Size (FWHM)    & $T_{\rm B}$     & $P$                        &  $L_{\rm R}$=$\nu P$   \\
($\mu$Jy)    & (GHz)     & north (mas) & east (mas)               & (mas)          & ($10^{8}$ K)    & ($10^{23}$ W Hz$^{-1}$)    &  ($10^{32}$ W)         \\
\hline                       
456$\pm$34   & 1.7       &  0               &  0                  &  1.25$\pm$0.05 &  1.90$\pm$0.44  & \hspace{2.5mm}2.9$\pm$0.2  & \hspace{2.5mm}4.8$\pm$0.3 \\
354$\pm$30   & 5         &  0               &  0                  &  0.45$\pm$0.02 &  1.26$\pm$0.20  & \hspace{2.5mm}2.3$\pm$0.2  & \hspace{0.8mm}11.4$\pm$0.9 \\
\hline                      
228$\pm$29   & 1.7       &  $-$375.6$\pm$0.1 &   235.8$\pm$0.1    &  2.03$\pm$0.18 &  0.28$\pm$0.09  &            $>$1.9$\pm$0.2  &            $>$3.2$\pm$0.4 \\
$<$$85$      & 5         &  ...             &  ...                &  $\la$8        &  ...            &               ...          &            ... \\
\hline   
\end{tabular}
\end{table*}

The high rest-frame brightness temperatures ($T_{\rm B}$$\sim$$10^{7}-10^{8}$ K) are indicative of a synchrotron origin of the radio emission in both sources. The brightness temperatures for star-forming galaxies typically do not exceed $\sim$$10^{5}$~K \citep{Cond92}. 
The 1.7-GHz radio powers of both compact individual VLBI-detected sources are at around $10^{23.5}$ W Hz$^{-1}$, the dividing value between starburst- and AGN-dominated sources \citep{Yun01}. The radio powers in the components of J1425+3231 exceed those in the most prominent nearby galaxies with nuclear starburst activity (Arp220, Arp299-A, Mrk273) whose powers are in the order of $10^{21}-10^{22}$ W Hz$^{-1}$ \citep[][and references therein]{Alex12}. This underpins our suspicion that 
compact radio jet emission from the immediate vicinity of a supermassive black hole is dominant in our VLBI-detected radio sources. A possible alternative explanation is, as we will discuss later, that at least one of the compact radio components marks a ``hot spot'', i.e. the location of a shock front in which a powerful relativistic jet interacts with the (inter)galactic medium much further away from the black hole.  

A comparison of the FIRST 1.4-GHz flux density (3.28~mJy), and the sum of the flux densities in our two components (0.68~mJy, at only a slightly different frequency of 1.7 GHz) indicates that there is significant extended radio emission, accounting for nearly 80 per cent of the total flux density. The extended emission is on $\sim$$0.1-1\arcsec$ angular scales, i.e. resolved out by our EVN observations but remains unresolved in the FIRST image ($\sim$$5\arcsec$ resolution). At 5~GHz, the analysis of the WSRT data obtained simultaneously with our EVN experiment gave a total flux density of 1.20$\pm$0.12~mJy. Comparing this with the VLBI component flux density (0.35~mJy; Table~\ref{modelfit}), we come to the same conclusion. In the latter case, variability as a potential cause of the difference in flux densities is excluded. The radio loudness of J1425+3231 as a whole (i.e. both components together), according to the definition by \cite{Kell89} as the ratio of 5-GHz radio and 4400-\AA\ optical flux densities, is $R$$\approx$9, which places the source near the dividing value of 10 between the radio-quiet and radio-loud objects.

The significant extended radio emission may suggest that, at least in part, active star formation is going on in the host galaxy. It qualitatively fits the picture of a galaxy merger containing two distinct AGN. Assuming that the difference between the 1.4-GHz FIRST and the 1.7-GHz EVN flux density (2.6~mJy) is entirely contributed by starbursts, we estimate the star formation rate (SFR) using the conversion relation between the SFR and the radio power \citep[e.g][]{Yun01,Hopk03}. This way we can place an upper limit, nearly 1000\,M$_\odot$yr$^{-1}$, for the SFR in the J1425+3231 system. Obviously this value is too high since there is no evidence for such an intensive star formation here. J1425+3231 is not known as a prominent far-infrared source, and the optical spectrum in \citet{Peng11} does not show a particularly strong [O II] emission, a potential indicator for high SFR. 
Therefore it is more plausible to assume that a dominant fraction of radio emission comes from extended jets or lobes which are resolved out by the EVN. The contribution of star formation is difficult to assess quantitatively.  

The radio spectra of the two compact sources in Fig.~\ref{Lband-image} are quite different. The NW object has a flat spectrum with $\alpha_1$=$-0.23$. (The spectral index $\alpha$ is defined as $S\propto\nu^{\alpha}$, where $S$ is the flux density and $\nu$ the frequency.) The value is characteristic of optically thick synchrotron jet emission from the nucleus. This is consistent with a Type 1 AGN as the primary component in this pair \citep{Peng11}.

The SE component is not detected at 5~GHz, therefore we can only derive an upper limit, $\alpha_2$$<$$-0.9$, for its two-point radio spectral index, considering 5$\sigma$ detection threshold.  
This spectral slope is consitent with AGN-related radio emission and does not preclude that this emission arises from the nucleus in a low-luminosity AGN. For example, both radio components of SDSS J1502+1115, another dual quasar with double-peaked narrow lines, have steep spectra, with spectral index $-0.80$ and $-0.92$ \citep{Fu11b}. 

The presence of arcsecond-scale extended radio emission in J1425+3231 raises the question if any of our compact VLBI-detected sources can be related to a hot spot. Such structures are seen in powerful FR-II \citep{Fana74} radio galaxies where the jets originating from the active nucleus propagate out to kpc distances, and form a termination shock by interacting with the ambient material. High-resolution VLBI imaging observations of radio galaxy hot spots are not very common \citep[see][for a review]{Ting08}, but include successful detections of the hot spots in nearby FR-II radio galaxies. Moreover, \citet{Gurv97} reported VLBI detection of a hot spot in a high-redshift radio galaxy (4C\,41.17, $z$=3.8) and interpreted the result as the deflection of the jet in a massive clump of the interstellar medium. While the measured brightness temperatures of the compact hot spot components in e.g. Pictor A \citep{Ting08} and Centaurus A \citep{Ting09} are only $\sim$10$^5$$-$10$^6$~K, the VLBI-detected component of 4C\,41.17 has nearly 10$^8$~K brightness temperature \citep{Gurv97}, comparable to the values we measured. 
Jet--cloud interactions are ubiquitous in e.g. compact steep-spectrum quasars. In case of 3C\,216, the interstellar matter deflects the initially straight jet and a compact, shocked region is observed on VLBI scales \citep[][and references therein]{Para00}.

Can then we rule out that one or both of the VLBI components of J1425+3231 are hot spots rather than compact AGN jets related to distinct black holes? The presence of at least one AGN is supported by multi-band (radio, optical, and X-ray) observational data, as will be discussed in Sect.~\ref{BH}.
In particular, the flat radio spectrum of the NW component suggests a radio ``core'' emission, while the steep spectrum of the SE component does not provide evidence for or against such an interpretation. 
But neither the available optical, nor the X-ray imaging data are sufficient to resolve two components with an angular separation of $\sim$440~mas. However, we are not restricted to the intrepretation of the radio data alone. \citet{Peng11} investigated the possibility that the double-peaked narrow O[III] emission lines in J1425+3231 are generated by effects other than a dual AGN: disk-like NLRs or biconical outflows. The latter could be qualitatively consistent with a jet termination shock scenario resulting in compact radio-emitting hot spots. \citet{Peng11} found that the relative widths of the red and blue emission line components, their different shifts with respect to the quasar's cosmological redshift, the consistent O[III] and H$\beta$ velocity offsets, and the high O[III] emission line luminosities comparable to that of an entire NLR, are hard to reconcile with the alternative scenarios.

More observational checks could be envisaged from the radio in the future. For example, a jet-like arsecond-scale radio structure connecting the compact VLBI sources would strongly suggest a relationship between the two. To reveal such a structure, deep imaging of J1425+3231 with intermediate ($\sim$50$-$100-mas) angular resolution would be needed. Higher-frequency (and thus higher-resolution) VLBI imaging could in principle prove if the sources are even more compact, ruling out the explanation involving hot spots. It is however demanding to reach a sensitivity sufficient to reliably image these weak, partly steep-spectrum sources at higher radio frequencies with VLBI.

\section{Exploring the dual supermassive black hole scenario}
\label{BH}

In this section, as a working hypothesis, we assume the validity of the dual AGN interpretation and investigate some physical properties of the system using multi-band data. 
J1425+3231 has been detected in X-rays as 1RXS J142506.7+323148 in the {\it ROSAT} All Sky Survey Faint Source Catalog \citep{Voge00}, and as CXOXB J142507.3+323137 with the {\it Chandra} X-ray Observatory in the XBootes survey as a point source at arcsecond resolution \citep{Kent05}. 
Even if our dual AGN is not resolved in X-rays, the observed relationship between the radio and X-ray luminosity and the mass of the central black hole \citep[e.g.][]{Merl03,Falc04} may be invoked, to give an estimate of the (total) black hole mass in this system. Indeed, by investigating a sample of X-ray emitting SDSS AGN, \citet{Li08} obtained a value of $\sim$$10^8$~M$_\odot$ for J1425+3231, using the FIRST 1.4-GHz radio luminosity and the {\it ROSAT} data. 
This is fully consistent with what is derived by \citet{Peng11} for the larger of the two suspected AGN, while the black hole mass estimated for the smaller one is two orders of magnitude less and thus negligible compared to the larger AGN. 
We note however that there may be large uncertainties in this estimate, because, as we have shown, not all the FIRST flux density is necessarily related to the AGN, and not all the low-luminosity AGN follow the \citet{Merl03} fundamental plane relation \citep[see e.g.][and references therein]{deGa11}. 

According to \citet{Peng11}, the [O III] $\lambda$5007 luminosity is $L_{\rm [O III]}$=3.2$\times$$10^{35}$~W for the blue system, i.e. the more massive of their two suspected AGN which we associate with our NW source due to its higher radio luminosity. Based on the results of \citet{Heck04}, the bolometric luminosity of this source is estimated as $L_{\rm bol}$=$3500 \, L_{\rm [O III]}$=1.1$\times$$10^{39}$~W. 
Assuming $10^8$~M$_\odot$ black hole mass for the primary object, the Eddington luminosity is $L_{\rm Edd}$=1.26$\times$$10^{31}$ $\frac{M_{\rm BH}}{\rm M_\odot}$~W=1.26$\times$$10^{39}$~W. The Eddington ratio, $L_{\rm bol}$/$L_{\rm Edd}$, is therefore close to unity. 
The mass accretion rate is 
\begin{equation}
\dot{m}=\frac{L_{\rm bol}}{\eta c^2}=\frac{L_{\rm bol}}{5.7 \times 10^{38} {\rm W}} \left( \frac{0.1}{\eta} \right) = 1.9\, \left( \frac{0.1}{\eta} \right) {\rm M}_\odot {\rm yr}^{-1}
\end{equation}
where $\eta$ is the mass-to-energy conversion efficiency. The high accretion rate and the measured radio luminosity clearly exclude the low-accretion model for the NW source in J1425+3231, such as the advection-dominated accretion flow (ADAF), and again suggests that the origin of the radio emission is from the jets \citep[e.g.][]{Wu05}. 

A similar calculation for the SE component \citep[$L_{\rm [O III]}$=2.1$\times$$10^{35}$~W, $M_{\rm BH}$=3.4$\times$$10^6$~M$_\odot$;][]{Peng11} results in a slightly lower bolometric luminosity, $L_{\rm bol}$=7.4$\times$$10^{38}$~W. The Eddington luminosity is $L_{\rm Edd}$=4.3$\times$$10^{37}$~W.
The corresponding Eddington ratio is 17, and the mass accretion rate is $\dot{m}$=1.3\,$(\frac{0.1}{\eta})$ M$_\odot$ yr$^{-1}$. 
This seems to support the finding of \citet{Peng11} that the less massive black hole (which we associate with the SE radio source) accretes at a super-Eddington rate. However, the Eddington ratio of 17 seems unrealistically high.

We note that the bolometric luminosity
is difficult to determine accurately. In practice, $L_{\rm bol}$ is estimated from monochromatic luminosities by applying different scaling factors \citep[e.g.][and references therein]{Runn12}. For the discussion above, we applied the average bolometric correction calculated for the [O III] emission line luminosity by \citet{Heck04}. However, the [O III] luminosity may only be an indirect estimator of the nuclear luminosity due to the unknown geometry and dust extinction of the NLR \citep[e.g.][]{Lama09}. The \citet{Heck04} relation did not correct the [O III] luminosity for dust extinction. It may result in an overestimate of the conversion factor from $L_{\rm [O III]}$ to $L_{\rm bol}$, and consequently the Eddington ratios and the mass accretion rates may be overestimated as well. Indeed, taking the bolometric corrections derived by \citet{Runn12}, and the measured continuum 5100-\AA\ luminosty $L_{5100}$=1.6$\times$$10^{33}$~W~\AA$^{-1}$ from \citet{Peng11} into account, we obtain a $\sim$25 times lower estimate for the the total bolometric luminosity of the entire source, i.e. both the NW and SE components of which the optical continuum emission is blended together. In this case, the data may be consistent with sub-Eddington accretion.

\section{Summary and conclusions}

We observed the double-peaked, narrow-line quasar J1425+3231 with the EVN to reveal its radio structure at a linear resolution of $\sim$50 pc. Based on compact radio emission detected from two distinct components with a projected linear separation of $\sim$2.6 kpc, we found that the object may contain a pair of AGN, as suspected from its optical spectrum by \citet{Peng11}. 
This separation is comparable to those of some other known dual AGN systems detected by radio interferometry \citep{Bond10, Fu11b}. 
Both radio sources in J1425+3231 have high brightness temperatures ($T_{\rm B}>10^7$K) typical for non-thermal synchrotron emission. They have high bolometric luminosities ($L_{\rm bol}$$\sim$$10^{38}$--$10^{39}$~W). The brightness temperatures, the bolometric luminosities, and the X-ray emission all support that the sources are powered by accretion onto SMBHs. J1425+3231 is therefore a good candidate for one of the very few cases of confirmed radio-emitting dual AGN. 
An alternative explanation that the less dominant compact radio source is a hot spot associated with a shock in a jet cannot be ruled out by the presently available data.
If the dual-AGN scenario is valid, the primary supermassive black hole is associated with the north-western radio source. 
The $\sim$$10^8$~M$_\odot$ mass of the larger AGN, derived from the comparison of radio and X-ray luminosities assuming the \citet{Merl03}
black hole fundamental plane relation \citep{Li08}, is consistent with the optical studies \citep{Peng11}.
Potential future observational verifications of the duality of J1425+3231 may include deep radio interferometric imaging at intermediate (sub-arcsecond) resolution, or adaptive optics imaging in the near infrared. The latter could possibly reveal signs of interaction between the merging host galaxies, such as a low surface brightness feature connecting the two nuclei \cite[e.g.][]{Rosa11}.

\citet{Dott12} note that high-resolution radio interferometry is not an efficient technique to {\it search} for rare objects as dual AGN. One of the reasons is the limited field of view. The other one is that majority of the AGN are not luminous in the radio. However, as shown by us and also by e.g. \citet{Bond10} and \citet{Fu11b}, sufficiently deep radio interferometric observations of a well-selected target sample could be an effective technique to {\it confirm} the existence of suspected AGN pairs. Radio interferometry allows us to spatially resolve two potentially distinct AGN and to determine some of their important geometric and physical properties.

\section*{Acknowledgments}

We are grateful to the chair of the EVN Program Committee, Tiziana Venturi, for granting us short exploratory e-VLBI observing time in 2010. 
We thank the anonymous referee for helpful suggestions that improved the paper.
The EVN is a joint facility of European, Chinese, South African, and other radio astronomy institutes funded by their national research councils. 
The WSRT is operated by the ASTRON (Netherlands Institute for Radio Astronomy) with support from the Netherlands Foundation for Scientific Research (NWO). 
e-VLBI research infrastructure in Europe is supported by the European Community's Seventh Framework Programme (FP7/2007-2013) under grant agreement RI-261525 NEXPReS.
This work was supported by FP7, Advanced Radio Astronomy in Europe, grant agreement 227290, the Hungarian Scientific Research Fund (OTKA, K72515), and the China--Hungary Collaboration and Exchange Programme by the International Cooperation Bureau of the Chinese Academy of Sciences (CAS). T. An thanks for the financial support by the Overseas Research Plan for CAS-Sponsored Scholars, the NWO, the National Natural Science Foundation of Science and Technology of China (2009CB24900), and the Science \& Technology Commission of Shanghai Municipality (06DZ22101). 
This research has made use of the NASA/IPAC Extragalactic Database (NED) which is operated by the Jet Propulsion Laboratory, California Institute of Technology, under contract with the National Aeronautics and Space Administration.

\label{lastpage}

\end{document}